\documentclass[aps,pra,twocolumn,amsmath,amssymb,superscriptaddress,floatfix]{revtex4-1}
\usepackage{amsmath}
\usepackage{amsfonts}
\usepackage{amssymb}
\usepackage{braket}
\usepackage{tikz-quantumgates}
\usepackage{tikzit}
\usepackage{url}          
\usepackage{graphicx}
\usepackage{xcolor}
\usepackage{makecell}
\usepackage{xspace}
\usepackage{algorithm}
\usepackage[noend]{algpseudocode}
\usepackage{quantikz}
\usepackage{balance}
\usepackage[unicode=true, colorlinks=true]{hyperref}

\begin{document}


\title{Evaluating low-depth quantum algorithms for time evolution on fermion-boson systems}



\author{Nathan Fitzpatrick}
\email[]{nathan.fitzpatrick@cambridgequantum.com}
\affiliation{Cambridge Quantum Computing Ltd, 9a Bridge Street, Cambridge, United Kingdom}

\author{Harriet Apel}
\affiliation{Cambridge Quantum Computing Ltd, 9a Bridge Street, Cambridge, United Kingdom}
\affiliation{Department of Physics and Astronomy, University College London, Gower Street, London, WC1E 6BT, United Kingdom}

\author{David Mu\~{n}oz Ramo}
\affiliation{Cambridge Quantum Computing Ltd, 9a Bridge Street, Cambridge, United Kingdom}



\begin{abstract}
Simulating time evolution of quantum systems is one of the most promising applications of quantum computing and also appears as a subroutine in many applications such as Green's function methods. In the current era of NISQ machines we assess the state of algorithms for simulating time dynamics with limited resources. We propose the Jaynes-Cummings model and extensions to it as useful toy models to investigate time evolution algorithms on near-term quantum computers. Using these simple models, direct Trotterisation of the time evolution operator produces deep circuits, requiring coherence times out of reach on current NISQ hardware. Therefore we test two alternative responses to this problem: variational compilation of the time evolution operator, and variational quantum simulation of the wavefunction ansatz. We demonstrate numerically to what extent these methods are successful in time evolving this system. The costs in terms of circuit depth and number of measurements are compared quantitatively, along with other drawbacks and advantages of each method. We find that computational requirements for both methods make them suitable for performing time evolution simulations of our models on NISQ hardware. Our results also indicate that variational quantum compilation produces more accurate results than variational quantum simulation, at the cost of a larger number of measurements.
\end{abstract}


\maketitle

\section{Introduction}
Quantum computing has emerged as a promising technology to study quantum dynamics problems, where the time evolution of a quantum system is essential to understanding many complex physical processes. There has been much interest in purely fermionic or bosonic dynamics on quantum computers, but little interest has been shown to fermion-boson systems which are central to physical processes such as electron-phonon interaction, quantum optics, and non-adiabatic dynamics in chemical reactions \cite{Polyak2019,Meyer2018,Spinlove2018,Wang2015,Richings2015,Norell2020}. To our knowledge, only a scarce number of works have been reported for the simulation of this interesting class of systems \cite{Macridin2018, Tacchino2021}.

The current quantum computing devices have small numbers of qubits with short coherence times due to limited control \cite{Bharti2021}. Therefore, Hamiltonian simulation frameworks that rely on many error corrected qubits such as Phase Estimation and Quantum Signal Processing are not currently possible  \cite{Lee2020,Kivlichan2020,Dong2020,Nam,Campbell2019,Childs2018,Babbush2018,Haah2018}. This situation has motivated the development of many time evolution algorithms for near-term quantum hardware. They can be classified into two broad families of algorithms: variational compilation methods and wavefunction ansatz variational methods. In the first case, the Trotterised time evolution operator is approximated by a simplified circuit obtained via a variational optimisation procedure. In the second case, time evolution is recast into an optimisation problem via an appropriate variational principle, leading to the construction of the wavefunction through an ansatz quantum circuit. Examples of the first family include Variational Fast Forwarding \cite{Cirstoiu2020,Gibbs2021}, Incremental Structured Learning (ISL)\cite{Jaderberg}, Variational Time Dependent Phase Estimation \cite{Klymko_vqpe}, circuit compilers for time dependent applications \cite{Bassman2020},  Adaptive Product Formula \cite{Zhang2020}, Quantum Imaginary Time Evolution \cite{Motta2020} and truncated Dyson Series \cite{Kieferova2019}. Examples of the second family of methods are Variational Quantum Simulation (VQS) \cite{Yuan,Mcardle2019,Mcardle2018,Lia}, Hardware-Efficient Real Time Evolution \cite{Benedetti2020}, Quantum Assisted Simulation \cite{Lau2021} and Truncated Taylor Series \cite{WeiZhongLau2021}.

 The time evolution protocols mentioned above have been predominantly demonstrated on purely fermionic systems like the hydrogen molecule or various model Hamiltonians. However, their relative performance remains to be understood, especially in the case of fermion-boson systems. As a particularly appropriate instance of this class of problems, we propose the Jaynes-Cummings Model (JCM) as a test case to benchmark the current NISQ quantum dynamics methods. The JCM describes two-level oscillators interacting with a quantised mode of an optical cavity in the semi-classical limit \cite{Jaynes1963}. This simple model can be represented with a two-qubit system and is a convenient toy model since it produces periodic oscillations in time. Extensions to the JCM that deal with an increasing number of oscillators encoded in extra qubits can be easily considered. We shall test these model systems with representative algorithms from the two families previously described, namely the ISL and VQS methods. We will compare the relative merits and flaws of the two methods at efficiently approximating  simulations done with direct Trotterisation of the time evolution operator.

The structure of this paper is as follows. Section \ref{sys_description} describes the theory of the Jaynes-Cummings model and its extensions, laying out details of the encoding used in the numerical simulations. Section \ref{time_methods} outlines the three algorithms used for simulating time evolution. Section \ref{num_results} presents the results for the three methods applied to the time evolution of our toy models, including a detailed cost comparison. The final section provides our conclusions regarding the relative performance of the time evolution algorithms investigated in this work for our models.

\section{System Description}
\label{sys_description}

Dipolar coupling between a quantised electromagnetic field mode and a two-level fermionic system is described by the quantum Rabi model \cite{Rabi1936, Rabi1937}. The two-level system (hereafter referred to as "the atom") could indeed be a natural atom, lending this model to atomic optics or instead an effective system arising in a solid state device. In general the model describes the simplest coherent coupling between a fermionic  system and a bosonic mode or photon. The quantum Rabi model Hamiltonian, $H_R$, is given by

\begin{equation}
H_R = \frac{\hbar\omega_c}{2}\hat{\sigma}_z + \hbar\omega_a \hat{b}^\dagger \hat{b} + \hbar \Omega \hat{\sigma}_x (\hat{b}+ \hat{b}^\dagger),
\end{equation}
where $\omega_c$, $\omega_a$ are resonance frequencies of the field mode and atom respectively; $\Omega$ quantifies the strength of the coupling between the field and the atom; $\hat{\sigma}_z$, $\hat{\sigma}_x$ are Pauli operators acting on the atom and $\hat{b}^\dagger$, $\hat{b}$ are the bosonic creation and annihilation operators.

\begin{figure}[]
\begin{center}
\includegraphics[width=\linewidth]{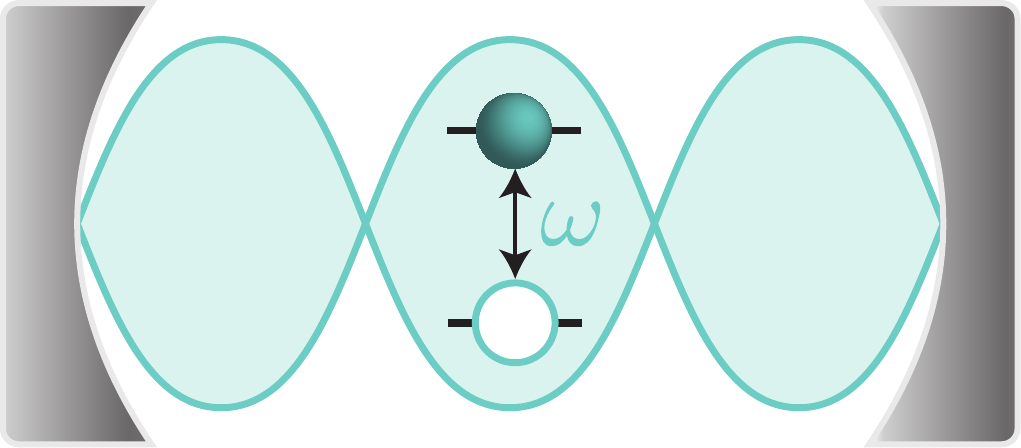}
\end{center}
\caption{The Jaynes-Cummings model of an atom coupled to a field mode in an optical cavity. Excitation or relaxation between the two states is induced by photon emission or absorption at frequency $\omega$ respectively.}
\label{fig_model}
\end{figure}

\subsection{Jaynes-Cummings Model}

In order to make the quantum Rabi model Hamiltonian tractable for quantum time evolution, various assumptions are then imposed. For ease, the system is assumed to be at resonance, $\omega_a = \omega_c$ (although we briefly explore off-resonant behaviour in Appendix \ref{off_res}). The JCM arises by performing the rotating wave approximation on the quantum Rabi model. A graphical representation of this model is shown in Figure \ref{fig_model}. Expressing the $\hat{\sigma}_x$ in terms of Pauli ladder operators ($\hat{\sigma}_x = \hat{\sigma}_+ + \hat{\sigma}_-$, where $\hat{\sigma}_\pm = \sigma_x \pm i \sigma_y$), in this regime it is assumed that there is strong coupling $\Omega/\omega<<1$ so the counter-rotating terms, $\hat{a}^\dagger \hat{\sigma}_+$ and $\hat{a}\hat{\sigma}_-$, contribute weakly to the dynamics and can be neglected. These approximations leave the Hamiltonian as:

\begin{equation}
H_{JC} = \hbar \omega (\hat{b}^\dagger \hat{b} + \frac{\hat{\sigma}_z}{2}) + \frac{\hbar \Omega}{2} (\hat{b} \hat{\sigma}_+ + \hat{b}^\dagger \hat{\sigma}_-) .
\end{equation}

The interacting term conserves the system's quanta through $U(1)$ symmetry and only exchanges between the coupled subsystems are allowed; it is also stationary in the interaction picture so the time dependence appears in the state. Denoting the excited and ground state of the bosonic mode and the atom as $\ket{e}/\ket{g}$ and $\ket{1}/\ket{0}$ respectively, consider a system initially in state $\ket{\psi(t=0)} = \ket{g,1}$. At time $t$ the state will evolve as:
 \begin{equation}
 \ket{\psi(t)} = \cos \left( \frac{\Omega t}{2}\right)\ket{g,1} - i \sin \left( \frac{\Omega t}{2}\right) \ket{e,0}.
 \end{equation}
Since the system is closed, the total energy will be constant in time. However, the probabilities of finding the system in a given eigenstate oscillate in time e.g. $\mathbf{P}_{g,1}(t) = \cos^2\left( \frac{\Omega t}{2}\right)$. This exchange of energy between the field and the atom can be used to observe the system's evolution in time by evaluating the expectation value of the field and atom Hamiltonian at each time step. 

\subsection{Tavis-Cummings Model}
An extension of the JCM comes from considering a single bosonic mode coupled with multiple ($N>1$) atoms; this model is known as the Dicke or Tavis-Cummings Model (TCM) \cite{Tavis1968}. Physically this setup could correspond to a higher dimensional spin or composite spin-$\frac{1}{2}$ system coupled to the cavity. Increasing the number of quanta in the model grows the Hilbert space exponentially, making it an ideal candidate for testing the complexity of the quantum computational simulation achievable on NISQ devices. The Hamiltonian for a system of $N$ atoms in the framework of this model is given by:
\begin{equation}
H_{TC} = \hbar \omega \hat{b}^\dagger \hat{b} +\hbar \omega \sum_{i=1}^N \frac{\hat{\sigma}_z^{i}}{2} - \frac{\hbar \Omega}{2} \sum_{i=1}^N (\hat{b}\hat{\sigma}_+^{i} + \hat{b}^\dagger \hat{\sigma}_-^{i}).
\end{equation}
As in the JCM case, this describes a closed system with a fixed number of quanta.

\subsection{Hamiltonian qubit encoding}

General Hamiltonians contain both fermionic and bosonic operators. Encodings translate Hamiltonians from the bosonic-fermionic representation into qubit representations. The Jordan-Wigner encoding is a common method to transform fermionic operators to quantum gates. However, as our numerical results realise time evolution in a atomic-photon system, bosonic operators appear in the Hamiltonian. Encoding bosonic operators for simulation on a quantum computer is less common than the fermionic case, so the procedure followed in order to obtain our qubit Hamiltonians is outlined here.

Using the Holstein–Primakoff transformation, we can relate spin operators to bosonic creation and annihilation operators \cite{Holstein1940}: 
\begin{equation}
\hat{b}^\dagger \hat{b} = S + \hat{S}_z, \qquad
\hat{b}^\dagger = \frac{\hat{S}_+}{\sqrt{S-\hat{S}_z}}, \qquad
\hat{b} = \frac{\hat{S}_-}{\sqrt{S-\hat{S}_z}} .
\end{equation}
The non-trivial form of these expressions is greatly simplified by making the semi-classical approximation, in which a truncated Taylor expansion in $1/S$ of these expressions is considered. This expansion is only valid for large $S$, so this approximation would only be strict with a large number of qubits representing the boson reservoir. However, this limit applied to the small qubit register in our models captures a significant part of the essential physics and is sufficient to demonstrate the time dynamics of the system. The simplified transform is given by:
\begin{equation}
\hat{b}^\dagger \approx \frac{\hat{S}_+}{\sqrt{2S}}, \qquad
 \hat{b} \approx \frac{\hat{S}_-}{\sqrt{2S}}.
\end{equation}
In our models, we have one single qubit representing our boson reservoir, so $S=1/2$ and the spin ladder operators can be easily translated into Pauli terms. 

This procedure is exemplified in the $N=1$ resonant JCM Hamiltonian with a two-qubit register, where the 0 superscript in the operators corresponds to the photon and the 1 superscript corresponds to the atom:
\begin{equation}
\begin{split}
H &=  \omega (\hat{b}^\dagger \hat{b} + \hat{\sigma}_z/2) + \frac{\Omega}{2} (\hat{b} \hat{\sigma}_+ + \hat{b}^\dagger \hat{\sigma}_-) \\
&= \omega (S + \frac{1}{2}\hat{\sigma}_z^0 + \frac{1}{2}\hat{\sigma}_z^1) + \frac{ \Omega}{4\sqrt{2S}} [(\hat{\sigma}_x^0 - i\hat{\sigma}_y^0)(\hat{\sigma}_x^1+i\hat{\sigma}_y^1) \\
& + (\hat{\sigma}_x^0 + i \hat{\sigma}_y^0)(\hat{\sigma}_x^1 - i\hat{\sigma}_y^1) ] \\
& =  \omega (S + \frac{1}{2}\hat{\sigma}_z^0 + \frac{1}{2}\hat{\sigma}_z^1) + \frac{\Omega}{4\sqrt{2S}} [2 \hat{\sigma}_x^0\hat{\sigma}_x^1 + 2 \hat{\sigma}_y^0 \hat{\sigma}_y^1 ].
\end{split}
\end{equation}

Encoding of the other Hamiltonians numerically simulated here follows similarly.

\section{time evolution methods}
\label{time_methods}

The Hamiltonian determines the time evolution of quantum states. For a system with a time independent Hamiltonian, $H$, an initial state, $\ket{\psi(t=0)}$, can be propagated by the unitary time evolution operator $e^{-iHt}$. Unitary operations can be performed on a quantum computer and, as previously explained, different methods have been developed to implement the time evolution operator on a quantum circuit. This section will briefly describe three different representative methods, later used to obtain numerical results. 

\subsection{Direct Trotterisation}

Qubit Hamiltonians acting in $(\mathbb{C}^2)^{\otimes n}$ can be represented as a linear combination of tensor products of Pauli operators acting on individual qubits: $H = \sum_k c_k \hat{P}_k^{(1)}\otimes\hat{P}_k^{(2)}...\otimes\hat{P}_k^{(n)}$. Direct exponentiation of this Hamiltonian is difficult, however quantum simulation algorithms avoid this using the Trotter asymptotic approximation:
\begin{equation}
\lim_{N\rightarrow \infty} \left( e^{iAt/N}e^{iBt/N}\right)^N = e^{i(A+B)t}.
\end{equation}
By performing a large but finite number of Trotter steps the unitary can be well approximated to arbitrary precision at the cost of circuit depth. Using this approximation the unitary time evolution operator becomes:
\begin{equation}
\hat{U}(t) \approx \left(\sum_k e^{-ic_k\hat{P}_k^{(1)}\otimes ... \otimes \hat{P}_K^{(n)}t/\hbar N} \right)^N.
\end{equation}
Exponentiations of this kind are easily implemented in the circuit model.

\subsection{Incremental structured learning}

Linearly increasing circuit depth with time and accuracy is a significant drawback of direct Trotterisation especially for near term applications. Benedetti \textit{et al.} \cite{Ostaszewski2021} developed circuit structure optimisation, later extended to the Incremental Structured Learning (ISL) technique developed by Jaderberg \textit{et al.},\cite{Jaderberg} which applies variational compilation ideas inspired by\cite{Jones,Khatri2019} to Trotterisation. In this technique, a so-called structural ansatz is built up to approximate each unitary created by adding a Trotter step to the time evolution operator. This optimisation is performed with respect to minimising the cost function 
\begin{equation}
C(\vec{\theta}) = 1 - |\bra{\bar{0}}\hat{Y}^\dagger(\vec{\theta})\hat{X}\ket{\bar{0}}|^2,
\end{equation}
where the target state is engineered to always be the vacuum state $\ket{\bar{0}} = \ket{0}^{\otimes n}$, $\hat{X}$ is the target unitary and $\hat{Y}$ is the approximate structural ansatz unitary built up incrementally. Adding the Trotter step to the previous structural ansatz creates the target circuit for the next time step. This method circumvents the issue of linearly increasing depth arising from direct Trotterisation by intelligently restructuring a hardware efficient circuit at each time step. However, this construction requires multiple evaluations of the cost function and introduces a classical optimisation subroutine. This methodology is described in more detail in Appendix \ref{isl_supp}.

\subsection{Variational Quantum Simulation}

Wavefunction ansatz methods follow a different approach to the methods previously discussed, and a good example is set out by the method of Li et al. \cite{Lia}, which uses a variational principle to simulate real time evolution. Unlike the variational compilation methods, this method requires a parameterised ansatz $\phi(\vec{\theta})$ to express the wavefunction of the system. This is an initial disadvantage since postulating a quantum circuit with sufficient freedom to represent the system at all times is non-trivial and often requires prior knowledge of the quantum system. For our numerical results, using a generalised or hardware-efficient ansatz was sufficient but this constraint could pose a challenge for general use of this technique. Instead of directly solving the Schr\"{o}dinger equation, McLachlan's variational principle is applied to represent the evolution of the wavefunction parameters with time. This evolution is given by the expression:
\begin{equation}
\sum_i A^R_{ij}\dot{\theta}_j = C^I_i,
\label{eq_vqs_diff}
\end{equation}
for time evolution governed by the Hamiltonian in its Pauli decomposition $H=\sum_k c_k\hat{P}_k$.
Given an ansatz with $n$ parameters, $\mathbf{A}^R$ is a $(n\times n)$ matrix and $\mathbf{C}^I$ is a vector of size $n$ defined as:
\begin{equation}
A^R_{ij} = \Re \left( \frac{\partial \bra{\phi(t)}}{\partial\theta_i}\frac{\partial \ket{\phi(t)}}{\partial\theta_j} \right)
\end{equation}
\begin{equation}
C^I_i =  \Im \left( \sum_k c_k \frac{\partial \bra{\phi(t)}}{\partial\theta_i} \hat{P}_k\ket{\phi(t)} \right).
\end{equation}
 We show the quantum circuits required to compute these objects for our models in Appendix \ref{vqs_supp}. The nature of this method requires multiple circuit evaluations to populate $\mathbf{A}^R$ and $\mathbf{C}^I$, whose size is heavily dependent on having an efficient ansatz with few parameters. However, the depth of these circuits is constant throughout the time evolution so trading numerous measurements for low depth circuits may be particularly advantageous in the NISQ era. Once these objects are computed, the parameters are advanced using an update method:
\begin{equation}
\vec{\theta}(t + \delta t) = \vec{\theta}(t)+\vec{\dot{\theta}}(t) \cdot   \delta t.
\end{equation}

A more detailed discussion of this algorithm's features is provided in Appendix \ref{vqs_supp}.

\section{Numerical results}
\label{num_results}

We present here our results for simulations of time evolution of the different models introduced previously with direct Trotterisation and the two variational algorithms previously described. The simulations have been run with our {\bf EUMEN} molecular simulation platform,\cite{eumen} interfaced with the {\bf tket} package for circuit optimisation \cite{Sivarajah_2020} and the {\bf QiskitAer}  shot based simulator \cite{Qiskit}. Parameters for our Hamiltonians have been set to $S = 1/2$, $\omega = 2$ and $\Omega = 10$.

\subsection{JCM results}

\begin{figure*}[]
\begin{center}
\includegraphics[trim= 75 60 35 49,clip,width=\textwidth]{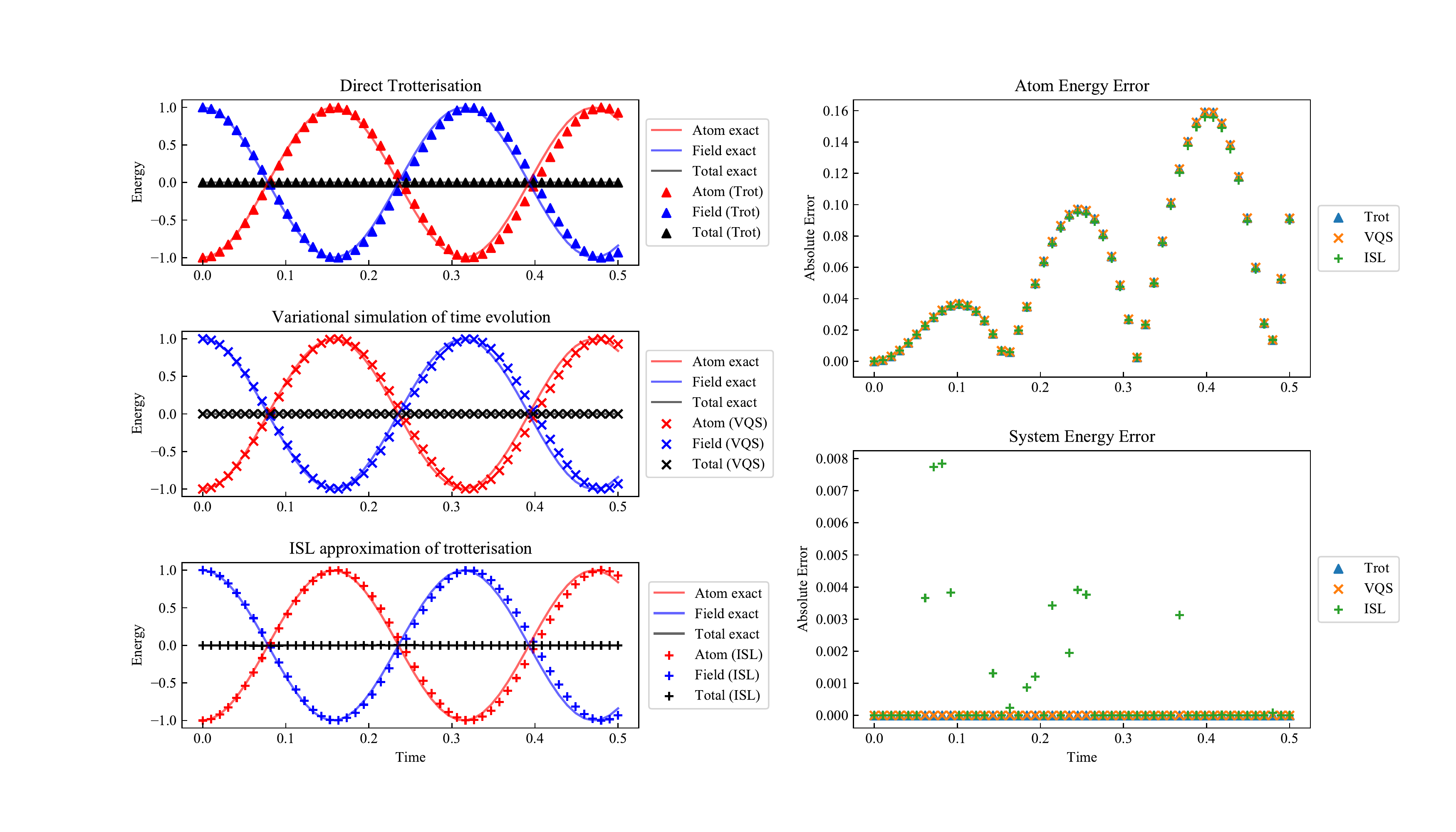}
\end{center}
\caption{\textit{Left:} Simulation of time evolution of the JCM system by three methods: direct Trotterisation, VQS and ISL. \textit{Right:} Absolute error in the dynamics with respect to time, for both the atom and the whole system. The methods simulate the time evolution to comparable accuracy, indicating  error is originating from discretisation for this model. }
\rule{\textwidth}{0.5pt}
\label{fig_jcm}
\end{figure*}

 The JCM on two qubits was simulated evolving in time using the three methods described above. Figure \ref{fig_jcm}  shows the energy evolution at each timestep for each method, along with a representation of the error in the energy estimation. All the methods agree qualitatively with the exact result calculated from the model using $\mathbf{P}_{g,1}(t) = \cos^2\left( \frac{\Omega t}{2}\right)$, validating the use of variational methods to study the JCM system. The error at this stage is dominated by Trotterisation error from discretising the system and so is similar for all three methods. This error can be decreased by increasing the number of Trotter steps and decreasing $\delta t$ in VQS. Since this discretising is causing a dephasing, it is intuitive that error is periodic like the exact function and is minimised where the function's gradient is minimum (at extrema of energies). When averaged over a period, the error is increasing with the number of time steps.

\subsection{TCM results}

\begin{figure*}[]
\begin{center}
\includegraphics[trim= 75 60 35 49,clip,width=\textwidth]{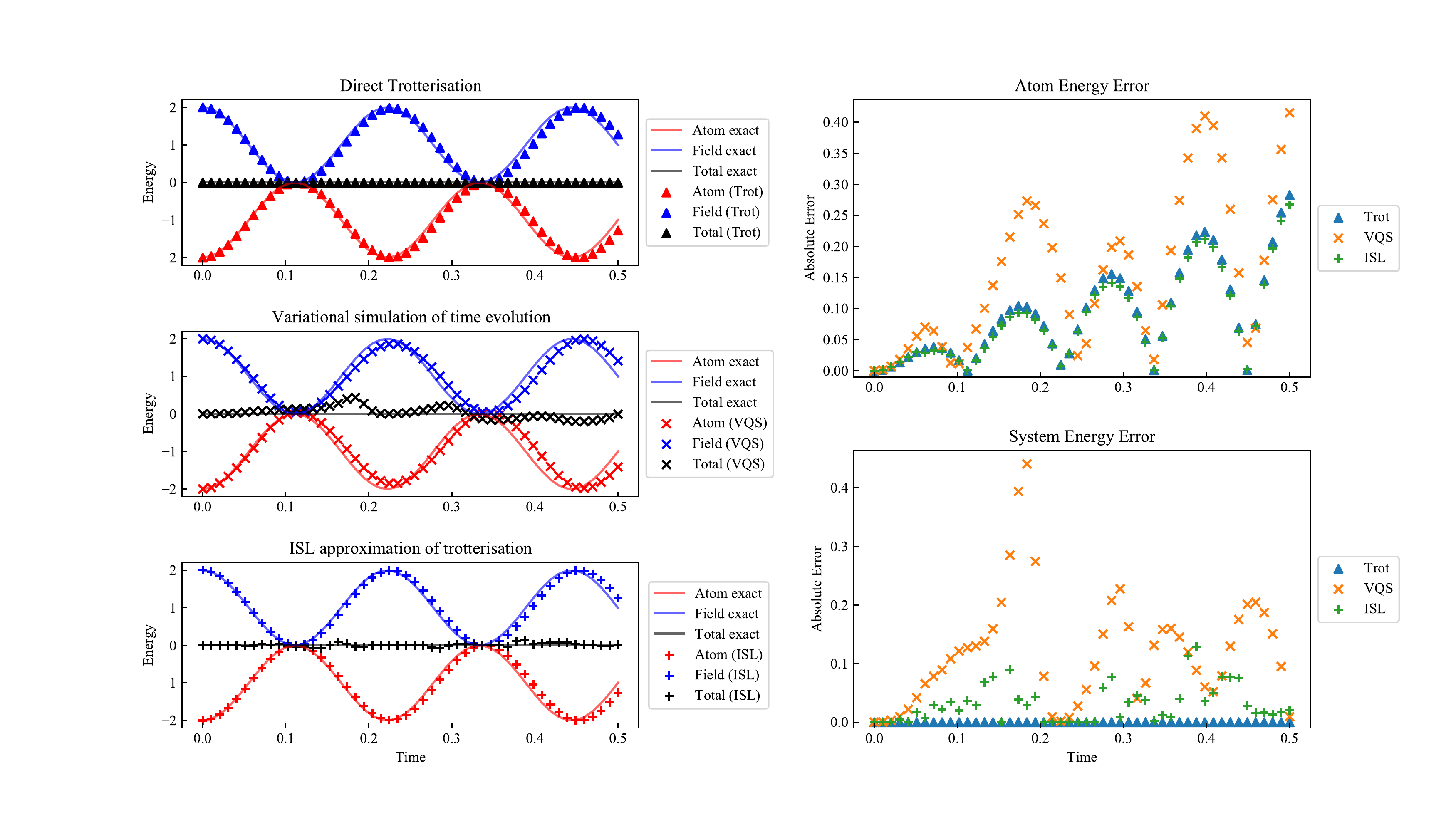}
\end{center}
\caption{\textit{Left:} Simulation of time evolution of the TCM system ($n=2$) by three methods: direct Trotterisation, VQS and ISL. \textit{Right:} Absolute error in the dynamics with respect to time. Unlike the JCM case, in this more complex system the methods' performance is varied with VQS being less accurate. }
\rule{\textwidth}{0.5pt}
\label{fig_tcm}
\end{figure*}

To demonstrate scaling, the TCM on three qubits (system with two quanta) was simulated evolving in time using the three methods described above. Results are displayed in Figure \ref{fig_tcm}. The different methods now produce varying accuracy and are compared with the analytical result calculated from the model using $\mathbf{P}_{g,1}(t) = \cos^2\left( \frac{\sqrt{n}\Omega t}{2}\right)$. As in the JCM case there is dephasing due to discretisation. However, VQS is now underperforming with a generally higher error that the other two methods. This could be due to the ansatz restricting the Hilbert space, demonstrating the need of careful consideration of the ansatz when using these methods. In contrast, ISL accuracy is comparable to direct Trotterisation. Increasing the number of layers in the ansatz may be a good solution to increase VQS accuracy. However, circuit depth and number of parameters to optimise also increase significantly. As current NISQ hardware can only manipulate a limited number of two-qubit gates before noise effects become untractable, this solution is currently unrealistic. In addition, the extra number of parameters may introduce additional complications to the ansatz optimisation process.

\subsection{Cost comparison}

\begin{figure}[h!]
\begin{center}
\includegraphics[width=\linewidth]{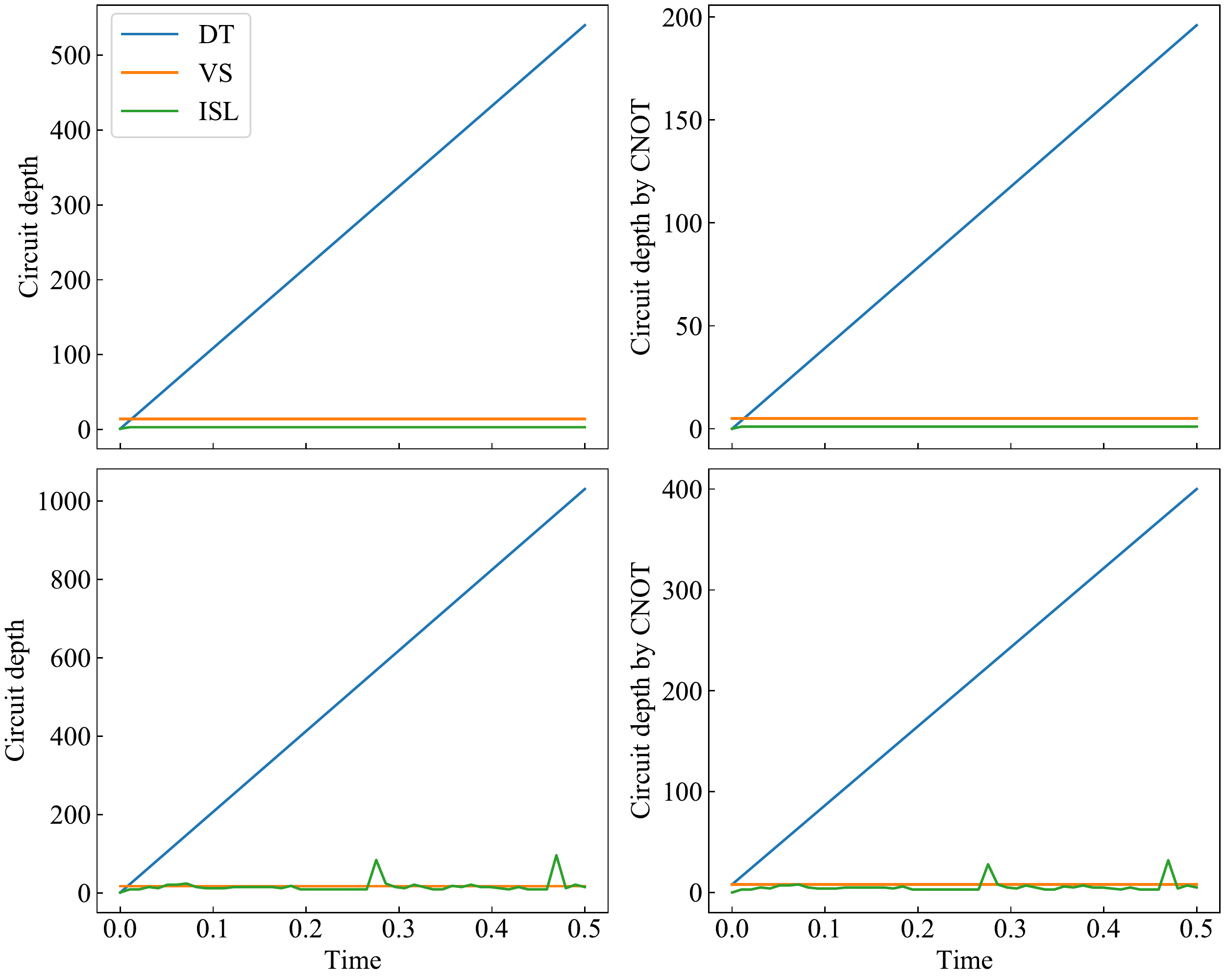}
\end{center}
\caption{\textit{Top row:} Circuit depth comparison for the JCM simulations. \textit{Bottom row:} Circuit depth comparison for the TCM simulations. In both cases both total circuit depth and 2-qubit gate depth are shown. Direct Trotterisation displays a linear depth dependence whereas  ISL and VQS show modest, almost constant, circuit depths.}
\label{fig_depth}
\end{figure}

\paragraph{Circuit depth} Due to the short decoherence times and high noise levels of current devices, deep circuits cannot be meaningfully measured. When simulating time evolution with all of the methods reported in this work, a discrete time step $\delta t$ is chosen, and a coarser discretisation results in a less accurate simulation. The number of time/Trotter steps is given by $m = T/\delta t$ where $T$ is the total evolution time. In all the results presented here $\delta t = 0.01$. Figure \ref{fig_depth} demonstrates how the circuit depths vary for each method in the JCM and TCM simulations. The linear increase with time steps for direct Trotterisation highlights the main disadvantage of this method. VQS has a constant depth determined by the choice of ansatz. ISL depth is also shown to be approximately constant and to predominately outperform VQS in creating shallow circuits with fewer two-qubit entangling gates. In this metric ISL and VQS clearly outperform direct Trotterisation demonstrating the motivation for their conception, however this advantage comes at other costs examined in the next section.

\paragraph{Number of circuit evaluations} Multiple shallow circuit evaluations, while feasible on NISQ hardware with manageable noise, generate large computational overheads. The more shots necessary to obtain the result, the more runtime is needed on the device, so it is desirable to minimise the number of circuit evaluations. Consider the scenario where one takes $k$ shots to produce an expectation value for a circuit, the system Hamiltonian is encoded as a sum of $n_p$ Pauli strings and the ansatz used for the VQS method contains $n_{pm}$ variable parameters. For ISL, the number of layers used (n$_l$) and the number of cost evaluations per layer (n$_{ev}$) must be considered as well, taking into account that n$_{ev}$ is dependent on the efficiency of the chosen optimisation strategy. The number of quantum circuit evaluations $c$ required per time step for each method scales as:
\begin{align}
&\text{DT:} \qquad c \approx kn_p \\
&\text{ISL:} \qquad c \approx kn_p + n_l \times n_{ev} \times k \\
&\text{VQS:} \qquad c \approx kn_p + \frac{1}{2}kn_{pm}(n_{pm}+1) + n_{pm}pk 
\end{align}

For the test case of the 3-qubit TCM Hamiltonian and a number of evaluations $k=1000$, we have $n_p = 7$, $n_{pm}=18$, average $n_{ev}=919$ and average $n_l=5.5$. The different $c$ values for each method can be found in Table~\ref{tab:table}. This test case gives an insight into the scale of measurements required for the different methods, however this cost is strongly model dependent. It is expected that the number of layers needed for ISL will increase as the number of qubits required by the model increases. Similarly, as larger systems are simulated, the ability to find an appropriate ansatz becomes less trivial and the number of variational parameters increases. How these two models compare for other larger models is uncertain, but in this test case  VQS performed better with fewer circuit evaluations. Variational simulation performs multiple measurements but over 3 fixed circuit structures with variational parameters, whereas ISL, by its nature, requires measurements of many different circuit structures. At implementation level this is a disadvantage of ISL.

\setlength{\tabcolsep}{12pt}
\begin{table*}[]
 \caption{Comparison of methods for time evolution on a quantum computer. The circuit evaluation is specific to the test case of TCM model with N=2, where $m$ is the number of time/Trotter steps.}
  \centering
  \begin{tabular}{lllll}\hline \hline
    Method    & Classical processing & Ansatz & Circuit Depth scaling    & Circuit evaluations/time step \\  \hline
    Trotter & None &No  & $\mathcal{O}(m)$  & 7000 \\
    ISL     & Optimisation & No  & $\mathcal{O}(1)$ & 5061500 \\
    VQS    &Tikhonov regularisation &  Yes  &$\mathcal{O}(1)$  &   304000\\
    \hline \hline
  \end{tabular}
  \label{tab:table}
\end{table*}

\paragraph{Ansatz} VQS is unique within our tested algorithms in that it requires a fixed ansatz to express the wavefunction.  The primary drawback of this feature is that the additional work required is specific to each system you want to simulate. This complication can be avoided by using hardware efficient ans\"{a}tze. However, it is not guaranteed that this type of ansatz will have sufficient degrees of freedom to describe the system as it evolves. However, an informed choice of ansatz that is physically motivated by the properties of the system may increase the robustness of the simulation; for example by limiting the associated Hilbert space to one which preserves a known symmetry.

\paragraph{Classical routines} Both ISL and VQS introduce classical subroutines. ISL requires classical optimisation since each layer includes variational parameters which are optimised to minimise the cost function. During our numerical simulations the simplistic Nelder-Mead method was used and more sophisticated algorithms, including conjugate gradient and BFGS, failed indicating barren plateaus on the cost function surface. However since in ISL the optimisation is iterative, there is freedom independent of system size to tailor the implementation of the algorithm to keep the number of variational parameters manageable, indicating that this method should scale efficiently. In contrast VQS avoids optimisation but instead requires the inversion of the matrix $\mathbf{A}^R$ in order to solve the differential equation \ref{eq_vqs_diff}. For our simulation, Tikhonov regularisation is used as this is robust when the matrix is close to singular. Unfortunately, the matrix size scales with the number of variational parameters in the chosen ansatz. Therefore, more complex systems requiring more qubits for their description may strain this subroutine.
The balance of these costs will determine which method is most suited to a given task on a given machine.

\section{Discussion and Conclusions}

Time evolution in the JCM and TCM was successfully simulated using the Holstein-Primakov tranform on a quantum computer, and the expected Rabi oscillations were successfully reproduced. The numerical results demonstrate that Trotterisation of the time evolution operator is an accurate method provided an appropriate time step is considered, although it is not friendly to current NISQ technology. Variational methods, either through circuit compilation or ansatz optimisation, are viable for simulating time dynamics on NISQ hardware. All of them can qualitatively, and in some cases quantitatively, describe features of these models. Despite the problems reported, these methods can be used to explore interesting physical phenomena appearing in both the JCM and TCM. We show in Appendix \ref{physics_app} some examples that illustrate this point.\newline

Our benchmarking uncovered no overall superior time evolution simulation algorithm: as expected, the best method will depend on the system of interest and resources available. Direct Trotterisation is the simplest to implement and for small systems and short times the circuit depth may be manageable. In the current regime of NISQ machines, the alternative methods lend themselves to longer time evolutions and provide another option when decoherence times prevent the deep circuits generated by Trotterisation from being realised. In this case the choice between VQS and ISL again lies in the particulars. If an suitable ansatz can be physically motivated with few parameters, VQS is efficient, requiring fewer circuit evaluations and with an attractively prescriptive methodology. However, ISL prevails in situations where the system is less understood and iteration to tolerance ensures the ansatz has sufficient freedom to accurately describe the unitary. There is also freedom within the ISL method regarding the optimisation protocol and construction of the approximate unitary that could be tailored to be efficient for the given problem.

Despite the simplifications introduced, computational requirements for the two variational methods are still challenging for current quantum computers. Further work is required to improve accuracy and reduce costs, including number of measurements and circuit depth, in order to make simulations of more complex systems feasible for NISQ devices.

\section*{Acknowledgements}
The authors thank Seyon Sivarajah for his helpful discussions on constructing pytket quantum circuits.

\balance

\bibliographystyle{apsrev4-1.bst}
\bibliography{JCMTCM.bib}



\clearpage

\onecolumngrid
\appendix

\section{Exploring model features}
\label{physics_app}

In our simulations we have focused on the basics of the model in order to compare the different methods for simulating time evolution. However, the Jaynes-Cummings model is interesting in its own right and by varying parameters, the aspects of the system's physics can be demonstrated using the time evolution methods described in this work. Obtaining the expected results is also an important sanity check that our computation is successfully simulating the physical system. 

\subsection{Off resonant behaviour}\label{off_res}

The main numerical results of this paper are developed for resonant Rabi oscillations. The cavity and oscillators can be detuned by adjusting the parameters of the toy model which generates an additional term in the Hamiltonian:

\begin{equation}
H = \hbar \omega_c (S + \frac{1}{2}\hat{\sigma}^0_z) + \frac{\hbar\omega_a}{2}\hat{\sigma}^1_z + \frac{\hbar(\omega_a - \omega_c)}{2}\hat{\sigma}^1_z + \frac{\hbar \Omega}{4\sqrt{2S}} \left[2 \hat{\sigma}^0_x\hat{\sigma}^1_x + 2 \hat{\sigma}^0_y \hat{\sigma}^1_y \right].
\end{equation}

\begin{figure}[!h]
\begin{center}
\includegraphics[trim={0cm 0cm 0cm 0cm},clip,scale=0.4]{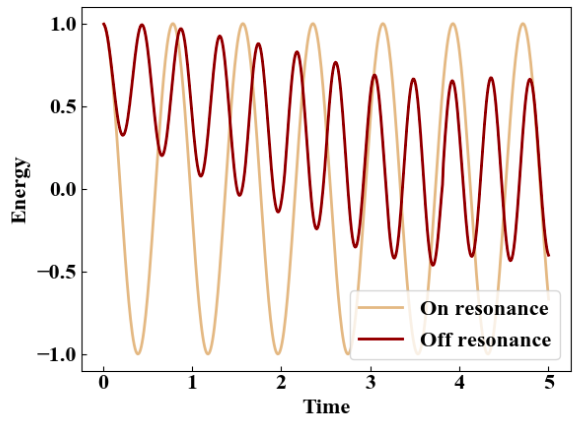}
\end{center}
\caption{Field energy evolution with time for the JCM at on and off resonance regimes.}
\label{jcm_detuning}
\end{figure}

The system, when evolved in time under this modified Hamiltonian, performed Rabi oscillations at a slightly different frequency and reduced amplitude. Figure \ref{jcm_detuning} demonstrates how detuning changes the state evolution by comparing the on and off resonance behaviour of the field energy. This behaviour is expected theoretically as detuning prevents the subsystem from transitioning between the two sublevels. 

\subsection{Investigating number of quanta}

\begin{figure}[!h]
\begin{center}
\includegraphics[trim={0cm 0cm 0cm 0cm},clip,scale=0.4]{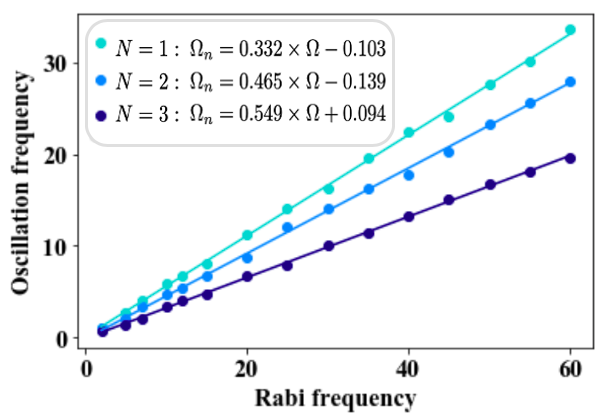}
\end{center}
\caption{Evolution of oscillation frequency with Rabi frequency calculated with trotterised time evolution.}
\label{fig_quanta}
\end{figure}

Our results focused on small systems of 2 or 3 qubits but the time evolution methods are all general to systems of any size. The toy model can also be adapted to demonstrate larger systems by adding quanta in the TCM which corresponds to adding qubits to the register.  Theoretically, the Rabi oscillation associated with an atom-field system of $n$ quanta will have frequency:
 \begin{equation}
 \Omega_n = \frac{\Omega \sqrt{n}}{2}.
\label{eq_quanta}
\end{equation} 
A manifestation of this relationship is seen numerically in Figure \ref{fig_quanta}, where the ratio of gradients agree approximately with what is expected from equation \ref{eq_quanta}:

 \begin{align}
 \frac{0.465}{0.332} &= 1.40 \approx \frac{\sqrt{2}}{\sqrt{1}} = 1.41  \\
  \frac{0.549}{0.332} &= 1.66 \approx \frac{\sqrt{3}}{\sqrt{1}} = 1.73.  
 \end{align} 
 
\section{ISL method}
\label{isl_supp}

We provide here a detailed overview of the ISL method, starting with the flow of the algorithm, then providing information about the optimisation steps and the qubit selection process implemented here. For a more in-depth explanation of the algorithm, we refer the reader to the work of Jaderberg et al. \cite{Jaderberg} whose method we follow closely. 

\subsection{Algorithm}

We start with the state of the system at time $t$; this state could be just the initialisation of the qubits or the circuit produced from the previous time step. The state is propagated in time by appending the usual Trotter step giving the target unitary $\hat{X}$. Instead of directly measuring properties of $\hat{X}$, an approximate unitary is built that describes with arbitrary accuracy the state of the system at time $t+\delta t$.\newline

This approximate unitary, $\hat{Y}$, is constructed by optimising a circuit with variational parameters $\vec{\theta}$. The optimisation is done with respect to minimising the cost function $C(\vec{\theta}) = 1 - |\bra{\bar{0}}\hat{Y}^\dagger(\vec{\theta})\hat{X}\ket{\bar{0}}|^2$, which has a minimum when $\hat{X}\hat{Y}^\dagger = \mathbb{I}$. The unitary $\hat{Y}$ is then obtained from $\hat{Y}^\dagger$. The low depth $\hat{Y}$ circuit is used to evaluate any quantities for the model (i.e. in this case the system, atom and field energies) before another Trotter step is appended and the system is repeated to continue propagating the state in time whilst avoiding the linear increase in depth. 

\subsection{Optimisation of the structural ansatz}

\begin{figure}[!h]
\begin{center}
\includegraphics[width = 4cm]{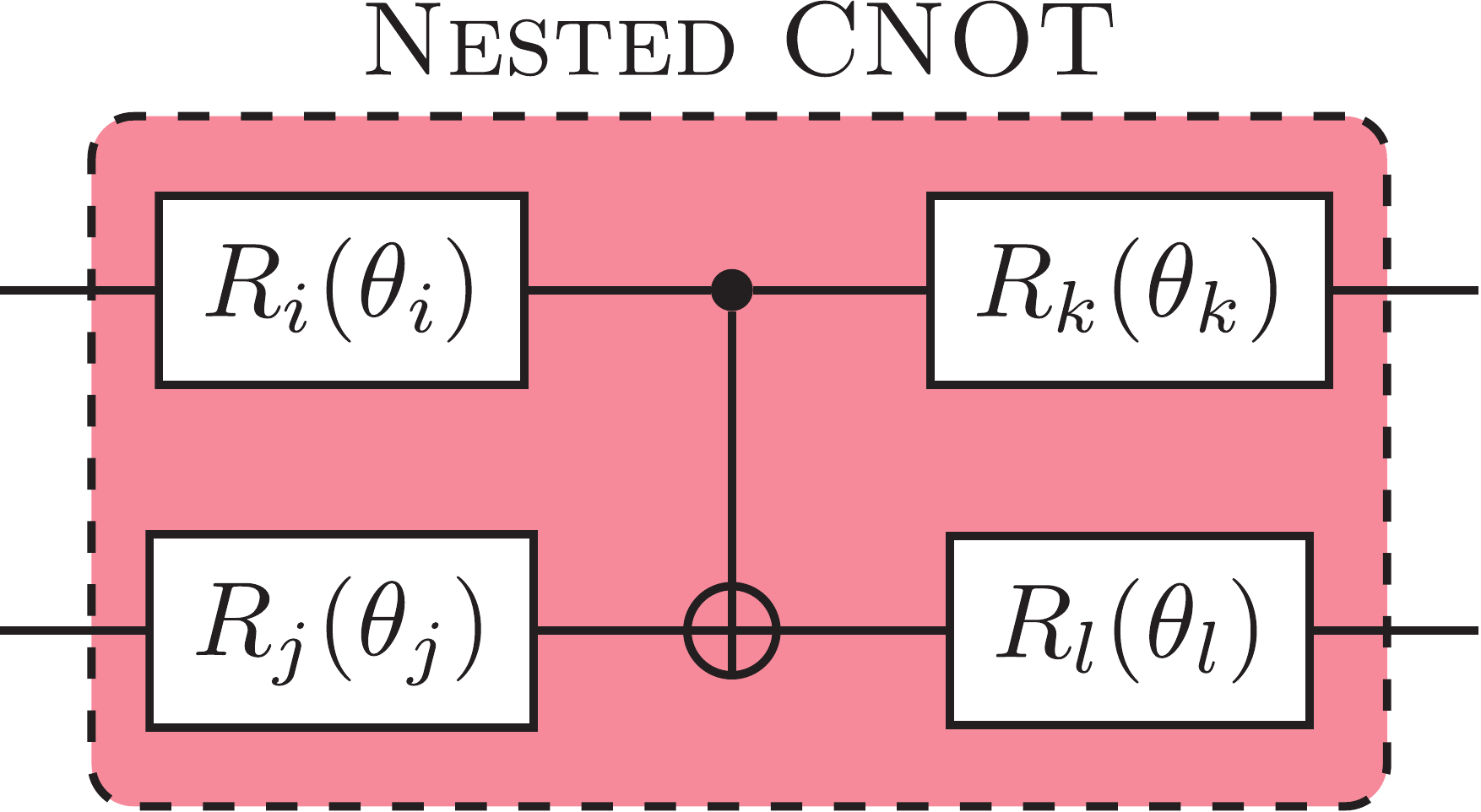}
\end{center}
\caption{Nested CNOT circuit block used to incrementally grow the ansatz to optimse $\hat{B}^\dagger$.}
\label{dressed_cnot}
\end{figure}

There is freedom in how $\hat{Y}^\dagger$ is constructed, which allows for flexibility depending on the available resources. A suitable approach is choosing a structural ansatz that is then optimised; this would greatly reduce the number of measurements required while introducing other issues. Choosing a suitable structural ansatz with enough freedom to describe the system is non-trivial for large systems, and the state's evolution quickly deviates from the true system dynamics if the approximated unitary is unable to drive the cost function below a strict tolerance value. To avoid these problems the ISL method constructs $\hat{Y}^\dagger$ by incrementally adding small circuit blocks. For our implementation the circuit block acts on two qubits and is a CNOT gate surrounded by single qubit rotation gates (see Figure \ref{dressed_cnot}). 

In order to minimise the circuit depth each layer is optimised carefully: 
\begin{enumerate}
\item two qubits are selected to which a new circuit block is appended;
\item in this circuit block the rotation angle and axis of three rotations is fixed while the fourth is optimised. This process is done sequentially for all four rotations in the block before the new cost function is evaluated. This step is then repeated until the improvement in cost between cycles is below certain threshold ({1\%} in this work);
\item fixing all rotation axes, all angles in the structural ansatz are optimised together before a final {\bf tket} circuit optimisation is applied to remove redundancies;
\item if the cost is below the pre-defined tolerance level, the cycle terminates. If it is above the tolerance level, the whole process is repeated.
\end{enumerate}

\begin{figure}[h!]
\begin{center}
\includegraphics[width=\textwidth]{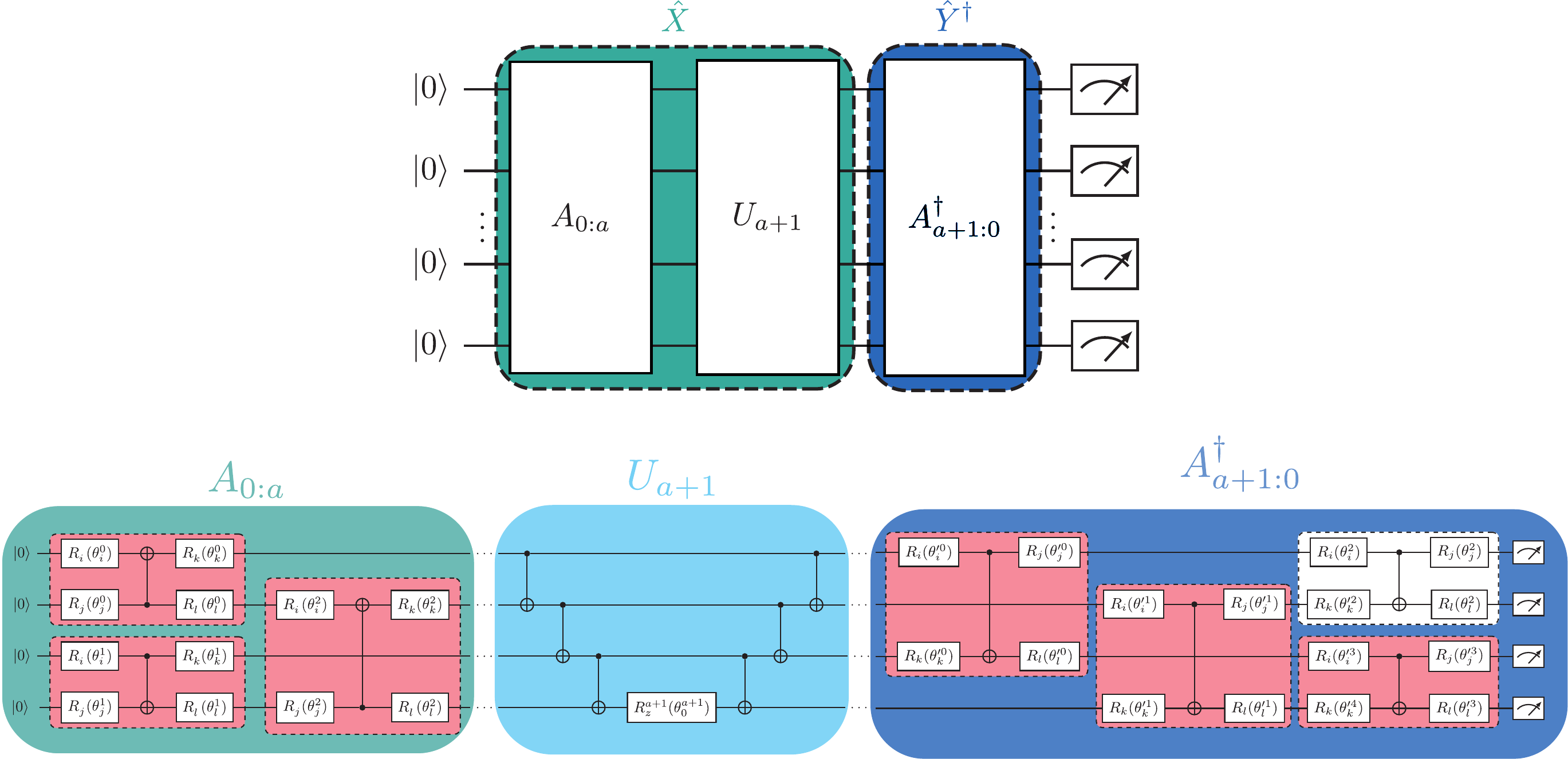}
\end{center}
\caption{A circuit description of the ISL update procedure evaluating the cost function $|\bra{0}\hat{Y}^\dagger(\vec{\theta})\hat{X}\ket{0}|^2$ for an individual trotter step denoted by the circuit $U_{a+1}$, where the intermediate compiled circuit is given by $A_{a+1:0}$. Nested CNOTs will continue to be appended until a pre-defined tolerance is reached.}
\label{compil_pic}
\end{figure}

A scheme of the compilation process is shown in Figure \ref{compil_pic}. For the numerical results a tolerance of $1\times 10^{-4}$ was used. 

\subsection{Choosing qubits}
Since the quantum system being simulated will, in general, require $n>2$ qubits to span the Hilbert space, careful placement of the 2 qubit circuit block at each augmentation of $\hat{Y}^\dagger$ has the potential to increase method performance. Selecting the pair of qubits with the highest entanglement outlined is the approached employed here. The entropy of formation of each pair of qubits is calculated by tracing out all other qubits and then calculating the entropy of formation, $E$. The qubit pair with the highest $E$ is selected conditional on it being a different pair to the previous circuit block.\newline

To calculate the entropy of formation of 2 qubits in a larger circuit the procedure in \cite{Ostaszewski2021} and \cite{Jaderberg} was followed. In summary:

\begin{enumerate}
\item the full density operator is obtained from the circuit;
\item all but the two qubits of interest are traced out leaving a ($4\times 4$) reduced density operator $\rho$;
\item the concurrence $C(\rho)$ is calculated
\begin{equation}
C(\rho) = \text{max} \{ 0, 2\lambda_\text{max} -\text{Tr} (R)\},
\end{equation}
where $\lambda_\text{max}$ is the largest eigenvalue of $\rho$. $R = \sqrt{\sqrt{\rho}\tilde{\rho}\sqrt{\rho}}$ and $\tilde{\rho}$ is the Hermitian conjugate of $\rho$ expressed in the magic basis:
\begin{equation}
\begin{split}
\ket{e_1} &= \frac{1}{2} (\ket{00} + \ket{11}), \qquad
\ket{e_2} = \frac{1}{2} (\ket{00} - \ket{11}),\\
\ket{e_3} &= \frac{1}{2} (\ket{01} + \ket{10}), \qquad
\ket{e_4} = \frac{1}{2} (\ket{01} - \ket{10});
\end{split}
\end{equation}
\item the entropy function is calculated as
\begin{equation}
E(p) = p \log_2(p) + (1-p) \log_2 (1-p)
\end{equation}
where $p(\rho)=\frac{1}{2} + \frac{1}{2} \sqrt{1-C(\rho)}$.
\end{enumerate}

\section{VQS method}
\label{vqs_supp}

In this appendix, we provide some additional background on the variational quantum simulation method. For full details, we refer to the original work in Refs. \cite{Yuan,Mcardle2019,Mcardle2018,Lia}. We first describe the method to recast the time evolution operator into a variational problem, and then we provide information about the wavefunction ansatz used for this method.

\subsection{Variational Principle}

We express the system's wavefunction as a series of unitary operators acting on the $\ket{\overline{0}}$ state:

\begin{equation}
\ket{\psi} =U_N...U_j U_{j-1}...U_1 \ket{\overline{0}} = V \ket{\overline{0}}
\end{equation}

If we allow $U_j$ to be a parametrised gate represented in terms of a Pauli generators in our state preparation ansatz

\begin{equation}
U_j = e^{-i\sum_a\theta_j\sigma_a/2}
\end{equation}

Using McLachlan's variational principle, \cite{Mclachlan1964} the real time evolution of the wavefunction is given by:
\begin{equation}
\sum_i A^R_{ij}\dot{\theta}_j = C^I_i
\end{equation}

After calculating $\mathbf{A}^R$ and $\mathbf{c}^I$ the parameters are advanced using an update method such as the Euler method:
\begin{equation}
\vec{\theta}(\tau + \delta \tau) = \vec{\theta}(\tau)+ \delta\tau \times  \vec{\dot{\theta}}(\tau)
\end{equation}

With this information  we can then derive a form of the real time evolution equations in terms of parameterised  unitaries. Given an ansatz with $n$ parameters $\vec{\theta}$, $\mathbf{A}^R$ is a $(n\times n)$ matrix and $\mathbf{C}^I$ is a vector size $n$ defined as:

\begin{equation}
\begin{split}
A^R_{ij}=\Re \left(\frac{\partial \bra{\psi}}{\partial \theta_i}\frac{\partial \ket{\psi}}{\partial \theta_j}\right)&=  \Re \left( \sum_{a,b} -\frac{i}{2} \cdot -\frac{i}{2} \bra{\overline{0}}U_1^\dagger ... U_{i-1}^\dagger \sigma_{a,i}^\dagger U_i^\dagger ... U^\dagger_N U_N...\sigma_{b,j} U_j U_{j-1}...U_1 \ket{\overline{0}}\right)\\
&= \Re \left(\sum_{k,l} - \frac{1}{4}\bra{\overline{0}}\tilde{V}_{k,i}^\dagger \tilde{V}_{l,j}\ket{\overline{0}}\right)
\end{split}
\end{equation}

\begin{equation}
\begin{split}
C^I_{i}= \Im \left( -\sum_k c_k \frac{\partial \bra{\phi}}{\partial\theta_i} \hat{P}_k \ket{\phi} \right)&= \Im \left( \sum_{k,a} -\frac{i}{2}c_{k}\bra{\overline{0}}U_1^\dagger ... U_{i-1}^\dagger \sigma_{a,i}^\dagger U_i^\dagger ... U^\dagger_N \hat{P}_k U_N...U_j U_{j-1}...U_1 \ket{\overline{0}}\right) \\
&= \Im \left(\sum_{k,a} f^*_{a,i}c_{k}\bra{\overline{0}}\tilde{V}_{a,i}^\dagger \hat{P}_{k} V \ket{\overline{0}}\right)
\end{split}
\end{equation}

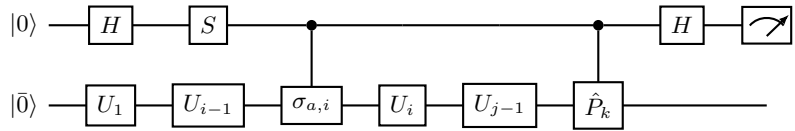
\begin{figure}[H]
\centering
\begin{quantikz}
\lstick{$\ket{0}$} & \gate{H} & \gate{S} & \ctrl{1} & \qw  & \qw &  \ctrl{1} & \gate{H}  & \meter{}   \\
\lstick{$\ket{\bar{0}}$} & \gate{U_1} &  \gate{U_{i-1}} & \gate{\sigma_{a,i}} &  \gate{U_i} & \gate{U_{j-1}} & \gate{\hat{P}_k} & \qw & \qw
\end{quantikz}
\caption{Quantum circuit for the measurement of the $C^I_i$ term in the JCM calculation.}
\label{fig:hadamard2}
\end{figure}

Where $\mathbf{C}^I$ is the imaginary part of the gradient vector with respect to wavefuction parameters and $\mathbf{A}^R$ is the real part of the metric tensor, which is related to the geometry of the parameter space. With $H=\sum_k c_k \hat{P}_k$ and $f_{a,i}$ as a phase factor. In order to calculate the $\mathbf{A}^R$ and $\mathbf{C}^I$ objects, we show the quantum circuits to calculate their elements $A^R_{i,j}$ and $C^I_i$ in Figures \ref{fig:hadamard} and \ref{fig:hadamard2}. The Hadamard test was chosen as the circuit primitive of choice but one could equally use another gradient protocol such as the phase shift method. 

\begin{figure}[H]
\centering
\begin{quantikz}
\lstick{$\ket{0}$} & \gate{H} & \qw & \ctrl{1} & \qw  & \qw &  \ctrl{1} & \gate{H}  & \meter{}   \\
\lstick{$\ket{\bar{0}}$} & \gate{U_1} &  \gate{U_{i-1}} & \gate{\sigma_{a,i}} &  \gate{U_i} & \gate{U_{j-1}} & \gate{\sigma_{b,j}} & \qw & \qw
\end{quantikz}
\caption{Quantum circuit for the measurement of the $A^R_{ij}$ term in the JCM calculation.}
\label{fig:hadamard}
\end{figure}

\subsection{Ansatz for numerical VQS simulations}

\begin{figure}[h!]
\centering
\begin{quantikz}
\lstick{$|0_1\rangle$} & \qw & \gate{R_x(\theta)} & \gate{R_y(\theta)}  & \gate{R_z(\theta)} & \ctrl{1} & \gate{R_x(\theta)} & \gate{R_y(\theta)}  & \gate{R_z(\theta)} & \ctrl{1} & \qw\\
\lstick{$|0_0\rangle$} & \gate{X} & \gate{R_x(\theta)} & \gate{R_y(\theta)}  & \gate{R_z(\theta)}& \targ{}  & \gate{R_x(\theta)} & \gate{R_y(\theta)}  & \gate{R_z(\theta)} & \targ{} & \qw \\
\end{quantikz}
\caption{Hardware-efficient ansatz used for the JCM in this work. General ansatz form with two layers. Two CNOT gates are needed to perform real time evolution with ansatz initialised to act as $U(\vec{\theta})=\mathbb{I}$. An $X$ gate is applied on the atom qubit to start the time evolution from the excited state.  }
\label{fig_ansatz_jcm}
\end{figure}
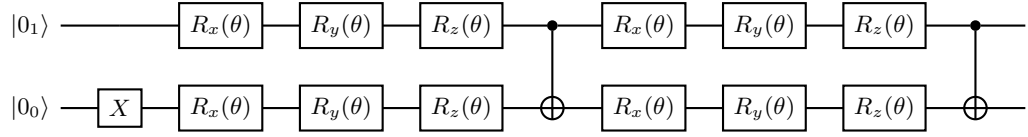

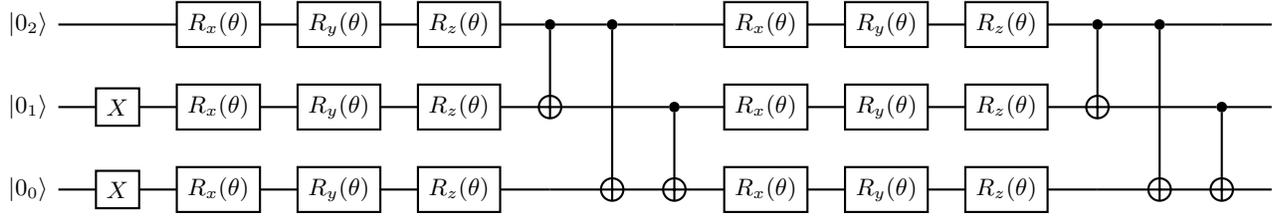
\begin{figure}[h!]
\centering
\begin{quantikz}
\lstick{$|0_2\rangle$} & \qw & \gate{R_x(\theta)} & \gate{R_y(\theta)}  & \gate{R_z(\theta)}& \ctrl{1} & \ctrl{2} & \qw & \gate{R_x(\theta)} & \gate{R_y(\theta)}  & \gate{R_z(\theta)}& \ctrl{1} & \ctrl{2} & \qw & \qw\\
\lstick{$|0_1\rangle$} & \gate{X} & \gate{R_x(\theta)} & \gate{R_y(\theta)}  & \gate{R_z(\theta)}& \targ{} & \qw & \ctrl{1} & \gate{R_x(\theta)} & \gate{R_y(\theta)}  & \gate{R_z(\theta)}&\targ{} & \qw & \ctrl{1} &  \qw\\
\lstick{$|0_0\rangle$} & \gate{X} & \gate{R_x(\theta)} & \gate{R_y(\theta)}  & \gate{R_z(\theta)}& \qw & \targ{} & \targ{}& \gate{R_x(\theta)} & \gate{R_y(\theta)}  & \gate{R_z(\theta)}& \qw & \targ{} & \targ{}& \qw\\
\end{quantikz}
\caption{Hardware-efficient ansatz used for TCM with $N=2$ in this work. General ansatz form with two layers. The ordering of the CNOTs in the entangling layers is specific to this setup (first qubit assigned to the field, second and third qubits assigned to the atom). The system is initialised with $X$ gates on the two atom qubits in order to introduce two quanta in the system.  }
\label{fig_ansatz_tcm}
\end{figure}

The VQS method requires an ansatz to express the model's wavefunction. Chosing an appropriate ansatz is non-trivial as it must have sufficient freedom through the variational parameters to access the system Hilbert space, while keeping a reasonable circuit depth. For the simulations presented in this work, hardware-efficient style ansatze where sufficient, with initialising $X$ gates so that the system begins with the oscillators in excited states and the field in the vacuum state at $t=0$. Figures \ref{fig_ansatz_jcm} and \ref{fig_ansatz_tcm} show the quantum circuits corresponding to the ansatze used in this work for the JCM and TCM, respectively.

\end{document}